\documentclass[a4paper,11pt]{article}

\usepackage[latin1]{inputenc}
\usepackage{authblk}
\usepackage{amsmath}
\usepackage{amsfonts}
\usepackage{amssymb}
\usepackage{graphicx}
\usepackage{appendix}
\usepackage{dcolumn}
\usepackage{caption}
\usepackage{subfigure}
\usepackage{float}
\usepackage{bm}
\usepackage{enumerate}
\usepackage{latexsym}
\usepackage{epsfig}
\usepackage[usenames]{color}
\usepackage[colorlinks=true]{hyperref}
\usepackage{pdfpages}
\title{Statistical Anisotropic  Gaussian Simulations of the CMB Temperature Field}
\author{Suvodip Mukherjee\footnote{suvodip@iucaa.ernet.in}\,\, and Tarun Souradeep\footnote{tarun@iucaa.ernet.in}\\
IUCAA, Post Bag 4, Ganeshkhind, Pune-411007, India\\}

\begin{document}

\maketitle{} 

\pagenumbering{arabic}
\thispagestyle{plain}
\markright{}

\begin{abstract}
Although theoretically expected to be Statistically Isotropic (SI), the observed Cosmic Microwave Background (CMB) temperature \& polarization field would exhibit SI violation  due to various  inevitable effects like weak lensing, Doppler boost and practical limitations of observations like non-circular beam, masking etc. However, presence of any SI violation beyond these effects may lead to a discovery of inherent cosmic  SI violation in the  CMB temperature \& polarization field. Recently, Planck presented strong evidence of SI violation as  a dipolar power asymmetry of the CMB temperature field in two hemispheres. Statistical studies of  SI violation effect require non-SI (nSI) Gaussian realizations of CMB temperature field. The nSI Gaussian temperature field  leads to non-zero off-diagonal terms  in the Spherical Harmonics (SH) space covariance matrix encoded in the coefficients of the Bipolar Spherical Harmonics (BipoSH) representation. We discuss an effective numerical algorithm, Code for Non-Isotropic Gaussian Sky (CoNIGS) to generate nSI realizations of Gaussian CMB temperature field of Planck like resolution with specific cases of SI violation.  Realizations of nSI CMB temperature field are obtained for non-zero quadrupolar $(L=2)$   BipoSH  measurements  by WMAP, dipolar asymmetry (resembles $L=1$ BipoSH coefficients) with a \emph{scale dependent} modulation field as measured by Planck  and for Doppler boosted CMB temperature field which also leads to $L=1$ BipoSH spectra. Our method, CoNIGS can incorporate any kind of SI violation and can produce nSI realizations efficiently.
%For SI violation such as Doppler boost, nSI realizations are made by incorporating the effect of aberration and modulation.
%can be incorporated without doing polynomial interpolation.
%temperature field have Gaussian random distribution with diagonal covariance matrix in harmonics space for Statistically Isotopic (SI) Universe. But violation of SI will lead to off- diagonal terms in the covariance matrix. WMAP and Planck has made a significant detection of violation of Statistical Isotropy of the CMB temperature field. Detections of  SI violation makes it necessary and interesting to prepare realizations of SI violated CMB temperature field.  Here we will discuss a computational technique developed by diagonalising the covariance matrix using Cholesky Decomposition to make SI violated realizations of temperature field. 

\end{abstract}

\section{Introduction}
Cosmic Microwave Background (CMB) is a very powerful probe of our Universe. Several important CMB experiments like COBE, WMAP, Planck, BOOMERanG, ACT, SPT etc., have opened an era of precision cosmology. Recent measurements from Planck \cite{Planck4} of the CMB temperature power spectrum matches well with the minimal $\Lambda$CDM model  at angular scales smaller than $2$ degrees. However, at  large angular scales, Planck \cite{Planck} has revealed possible signature of Statistical Isotropy (SI) violation of CMB temperature field, which is beyond our present understanding. SI violation can be measured by Bipolar Spherical Harmonics (BipoSH) coefficients, introduced in CMB temperature measurements by  Hajian \& Souradeep \cite{ts,tsa}.  These are the linear combinations of off-diagonal terms in the covariance matrix and arise from the correlation between different CMB multipoles $l$,  in Spherical Harmonics (SH) space. \\
Violation of SI is an inevitable consequence of known effects such as  weak lensing, Doppler boost due to our motion with respect to the CMB rest frame \cite{anthony, Planck_dop, Suvodip}. SI violation is also expected from theoretically motivated possibilities such as non-trivial cosmic topology \cite{nsi_topo}. Presence of systematics in experiments like non-circular beam, masking etc., can also lead to SI violation.  Detection of non-SI (nSI) signal that are not consistent with SI Cosmological models, can indicate a breakdown of Cosmological Principle (CP). Recently, WMAP \cite{ben} and Planck \cite{Planck} measured a significant detection of non-zero BipoSH coefficients. A viable solution for WMAP's non-zero measurement of BipoSH  is due to non-circularity of WMAP beam is shown by several authors \cite{ben, beam1, beam2, beam3, beam}. However, the origins of the BipoSH measurement from Planck is beyond the understanding of present SI Cosmological models. Thus, quest for measuring  nSI signal are of primary interest in understanding our Cosmological model, and also to remove the effect of systematics on CMB maps. \\ 
To investigate these effects, it is important  to generate large number of  simulations of  nSI Gaussian CMB maps of temperature and polarization field. These simulations help in determining the statistical properties of nSI sky due to the physical process mentioned above and the ability of different estimators on sky maps to discern \& distinguish between them.  Various algorithms are available to incorporate SI violation due to weak lensing of CMB temperature and polarization maps \cite{lenpix, hirata, sudeep,basak}. 
Non-Gaussian (NG) SI maps were made for studying non-Gaussianity in CMB sky by Liguori et al. \cite{ng1}, Regan et al. \cite{regan} and Rocha et al. \cite{ng}. Statistical anisotropic maps including  a specific case like Doppler boost signal are made by Chluba \cite{chulba} and Catena \& Notari \cite{catena}. \\
In this paper we present  an efficient numerical algorithm, Code for Non-Isotropic Gaussian Sky (CoNIGS), to make Monte-Carlo realizations of  nSI CMB temperature field. In this algorithm, we efficiently Cholesky decompose the covariance matrix in SH space. A key feature of our approach is that it can make simulations for any type of SI violation present in CMB temperature field.\\
The paper is organized as follows, in Sec. 2, we review some of the existing topics, we use in this paper. In Sec. 3, we discuss the method to make nSI Gaussian maps of temperature. In Sec. 4, we discuss the results from the simulated maps for three different cases. Discussion and conclusions are given in Sec. 5.

\section{Review of statistical properties of CMB temperature field}
In our present understanding, temperature fluctuation of CMB  originated from the quantum fluctuations of the ground state of the single inflation scalar field. This implies that the statistics of the CMB temperature fluctuation is expected to be Gaussian with zero mean. Recent results from experiments like Planck \cite{Planck_ng} also place fairly strong constraints on primordial non-Gaussianity to be consistent with zero. In the next sections, we briefly review the statistics of CMB temperature field and its covariance matrix.
\subsection{CMB temperature field}
Temperature anisotropy of CMB  sky map $\Delta T (\hat n)$ can be expanded in the orthonormal space of Spherical Harmonics (SH) functions on the sphere,
\begin{align}\label{eqbi1}
\Delta T(\hat n)= \sum_{lm} a_{lm}Y_{lm}(\hat n),
\end{align}
where, $a_{lm}$ are the coefficients of Gaussian random field in SH space. The statistics of  any Gaussian distribution can be specified by its  covariance matrix, $G_{ij}\equiv \langle a^*_{j}a_{i}\rangle$ where, $i=l(l+1)+m-3$ is a single positive index representing an SH multipole  $(l,m)$. The probability distribution of the temperature field is given by,
\begin{align} \label{eq1}
P[{a_{j}}]= \frac{1}{(2\pi)^{N/2}\sqrt{|G|}}\exp{[-\frac{(a^{\dagger}_{i} G^{-1}_{ij}a_{j})}{2}]},
\end{align}
where, we assume the  Einstein summation rule. Given the largest multipole $l_{max}$, the covariance matrix is of size 

\begin{align}\label{eq3n}
N&= l_{max}(l_{max}+1)+l_{max}-3.
\end{align}

Under the assumption of SI temperature field, the two-point correlation function on the sphere (sky) depends only upon angular separation between the two-points. This implies a diagonal covariance matrix $G$, given by the angular power spectrum $C_l$ by,
\begin{align} \label{eq2a}
G_{ij} \equiv \langle a^{*}_{j}a_{i}\rangle = C_l \delta_{ij}.
\end{align}
But for SI violated maps, covariance matrix is not diagonal. The non-zero off-diagonal elements of non-SI (nSI) temperature field can be expressed by Bipolar Spherical Harmonics (BipoSH) coefficients introduced by Hajian \& Souradeep \cite{ts}. In the next section we review the BipoSH coefficients and its connection to the covariance matrix.

\subsection{Formalism of BipoSH coefficients for CMB temperature field}
As mentioned in the previous section,  the CMB temperature fluctuations are Gaussian random field with zero mean and we can express the statistics of this field by the two-point correlation function. In the full generality, the two-point correlation of SH coefficients of the CMB temperature anisotropy $\langle a^*_{j} a_{i}\rangle$ can be expanded in the tensor product basis of two SH space as, 
 \begin{align}\label{eqbi2}
\langle a^{*}_{j} a_{i} \rangle = \sum_{LM} A^{LM}_{ll'}(-1)^{m'}C^{LM}_{lml'-m'} ,
\end{align}
where, $A^{LM}_{ll'}$ are called the BipoSH coefficients \cite{ts} and $C^{LM}_{lml'm'}$ are the Clebsch-Gordan (CG) coefficients. These BipoSH coefficients with $L=0, M=0$, are the diagonal elements related to the angular power spectrum by $A^{00}_{ll'} \delta_{ll'}= (-1)^l C_l \sqrt{(2l+1)}$.\\ For nSI case, BipoSH coefficients are non-zero for $L \neq 0, M\neq 0$. We can relate eq.\eqref{eqbi2} to the covariance matrix $G$ by,
 \begin{align}\label{eqbi3}
 \begin{split}
G_{ij} \equiv \langle a^{*}_{j} a_{i}\rangle &= \sum_{LM} A^{LM}_{ll'}(-1)^{m'}C^{LM}_{lml'-m'} ,\\
G_{ij} \equiv \langle a^{*}_{j} a_{i}\rangle &= C_l \delta_{ij} + \sum_{LM; L\neq 0} A^{LM}_{ll'}(-1)^{m'}C^{LM}_{lml'-m'},
\end{split}
\end{align}
where  $i$ and $j$ are related with $l, m$ and $l', m'$ by
\begin{align} \label{eq3a}
\begin{split}
i&= l(l+1)+m-3,\\
j&= l'(l'+1) +m'-3.
\end{split}
\end{align}
Using the definition of BipoSH spectra from WMAP \cite{ben}, eq.\eqref{eqbi3} becomes,
\begin{align}\label{eqbi3c}
 \begin{split}
%G_{ij} \equiv \langle a^*_{i} a_{j}\rangle &= \sum_{LM} A^{LM}_{ll'}(-1)^{m'}C^{LM}_{lml'-m'} ,\\
G_{ij} = C_l \delta_{ij} + \sum_{LM; L\neq 0} (-1)^{m'}\alpha^{LM}_{ll'} \frac{\Pi_{ll'}}{\Pi_{L}}C_{l0l'0}^{L0}C^{LM}_{lml'-m'},
\end{split}
\end{align}
where, BipoSH coefficients $A_{ll'}^{LM}$, are related with BipoSH spectra, $\alpha^{LM}_{ll'}$, by\\
\begin{align} \label{eq3b}
A^{LM}_{ll'}= \alpha^{LM}_{ll'} \frac{\Pi_{ll'}}{\Pi_{L}}C_{l0l'0}^{L0},
\end{align}
and following the notation in \cite{varsha}
\begin{align} 
\Pi_{l_1l_2 \ldots l_n}&= \sqrt{(2l_1+1) (2l_2+1)\ldots (2l_n+1)}.
\end{align}
Any non-zero measurement of BipoSH coefficients indicates violation of SI of our Universe. Weak lensing, Doppler boost, masking and non-circular beam are some of important effects that lead to SI violation in the observed  CMB sky even when the underlying CMB signal is SI. 
 
\section{Method: Simulations of nSI CMB temperature field}
%Violation of SI is the inevitable effect in weak lensing, Doppler boost, topology of the Universe etc. Presence of systematics like non- circular beam, masking can also lead to violation of SI.  Detection of any nSI signal, can result into assumption of SI Universe. Recently, WMAP \cite{ben} and Planck \cite{Planck} measured a significant detection of non-zero BipoSH coefficients ($A^{LM}_{l_1l_2}$) for $L=2$ and $L=1$ respectively. A viable solution for WMAP's non- zero measurement of BipoSH  is due to non- circularity of WMAP beam as shown by Nidhi et al. \cite{beam},  however the origin of the BipoSH measurements from Planck are beyond our understanding of SI Cosmological models. Thus test for nSI effect are of primary importance for understanding our Universe, and also to remove the systematics. \\
The central difference between the covariance matrix in SH space for SI and nSI CMB temperature field is the presence of off-diagonal terms in the SH space covariance matrix. This implies that different modes $(l,m)$ of CMB are not independent in SH space for nSI temperature field. The key idea we implement is to make a change of basis from SH space to another space in which this covariance matrix is diagonal. A linear transformation does not change the Gaussian statistics of the field. In the new space, different modes are no more correlated, but realizations of CMB temperature in this basis are manifestly SI violated. On performing an inverse transformation of CMB temperature map from the new space to SH space, gives us the nSI CMB temperature maps in SH space. 
\subsection{Cholesky decomposition  of the covariance matrix} \label{cdbasic}
In this section, we discuss the method of diagonalization of the covariance matrix by Cholesky Decomposition (CD) algorithm. In CMB analysis,  CD has been implemented by Gorski \cite{gorski_2} for diagonalization of coupling matrix for SH on the cut sky.
 When SI violation is captured in  a limited set of non-zero BipoSH coefficients (like $L=1$ captures Doppler boosted sky, scale dependent dipole modulation, $L=2$  captures quadrupolar anisotropy and any other pattern), it is possible to implement an efficient code that scale by $~N^{0.853}$ implying  $\sim l_{max}^{1.70}$ ($l_{max}$ determines the angular resolution of the map), instead of the well-known $N^3/6$ scaling of the CD algorithm in general cases. 
 The covariance matrix $G$ of a Gaussian distribution is always positive definite which satisfies the following two conditions \cite{golub},
\begin{enumerate} 
\item 
For any vector  $y \in \mathbb{R}, y^TGy>0$. This condition implies a strong constrain on diagonal terms, i.e. $G_{ii}>0$.

\item

The determinant of the covariance matrix should be positive, i.e., $|G| \geq 0$. Since the inverse of the covariance  matrix $(G^{-1})$ must exist, this makes $|G|>0$.\\
\end{enumerate}
Also because of the dependence of the elements of covariance matrix on CG coefficients as mentioned in eq.\eqref{eqbi3c}, the elements of covariance matrix also obey the properties of CG coefficients \cite{varsha},
%\begin{enumerate} 
%\item 
\begin{align} \label{eq6}
\begin{split}
C^{LM}_{lml'm'} &\neq 0; \hspace{2cm} \text{iff} \, \, \,|l-l'|<L<l+l';\,\,\,
m+m'= M;\\
%\end{split}
%\end{align}
%\item
%\begin{align} \label{eq6a}
%\begin{split}
C^{L0}_{l0l'0} &\neq 0; \hspace{2cm} \text{iff} \, \, \, l+l'+L= 2n;\,\, \text{n is an integer}.
\end{split}
\end{align}
%\end{enumerate}
Due to the conditions mentioned in eq.\eqref{eq6}, the covariance matrix is sparsely populated, which makes the diagonalization of the covariance matrix efficient.
%\subsubsection{Effect of cholesky decomposition of CMB temperature map}
Cholesky decomposition  leads to decomposition of covariance matrix into a lower triangular matrix $\mathcal{L}$ and  its conjugate transpose $\mathcal{L}^\dagger$, 
\begin{align} \label{eq4}
G= \mathcal{LL}^\dagger, 
\end{align}
where elements of $\mathcal{L}$ are related to the elements of $G$ by \cite{golub},
\begin{align} \label{eq4a}
\begin{split}
\mathcal{L}_{ii}&= \sqrt{(G_{ii}- \sum_{k=1}^{i-1}\mathcal{L}^2_{ik})}\,,\\
\mathcal{L}_{ji}&= \frac{(G_{ij}- \sum_{k=1}^{i-1}\mathcal{L}_{ik}\mathcal{L}_{jk})}{\mathcal{L}_{ii}}; \hspace{1cm}   j= i+1,.....n .
\end{split}
\end{align}
On performing CD on covariance matrix $G$, the modification to the probability distribution function eq.\eqref{eq1} can be written as,
\begin{align} \label{eq1d}
\begin{split}
P[{a}]= &\frac{1}{(2\pi)^{N/2}\sqrt{|G|}}\exp{[-\frac{(a^{\dagger} (\mathcal{L}\mathcal{L}^\dagger)^{-1}a)}{2}]},\\
=&\frac{1}{(2\pi)^{N/2}\sqrt{|G|}}\exp{[-\frac{(a^\dagger(\mathcal{L}^\dagger)^{-1}\mathcal{L}^{-1}a)}{2}]},\\
=&\frac{1}{(2\pi)^{N/2}\sqrt{|G|}}\exp{[-\frac{(\mathcal{L}^{-1}a)^{\dagger}(\mathcal{L}^{-1}a)}{2}]},\\
= &\frac{1}{(2\pi)^{N/2}\sqrt{|G|}}\exp{[-\frac{x^\dagger Ix}{2}]},
\end{split}
\end{align}
where, we define $x_j=\mathcal{L}_{ji}^{-1}a_{i}$. $x$ is the Gaussian CMB temperature map in the new space with unit variance. On performing inverse of this transformation, we get, $a_{i}= \mathcal{L}_{ij}x_j$. This map $a_i$, is manifestly SI violated in SH space and  average over random realizations of $a_i$ should match the  input BipoSH coefficients.
%In the next section, we discuss the algorithm of making simulation of CMB temperature field.
%Using the property of positive definiteness of the covariance matrix, discussed in the previous section, we can diagonalize the covariance matrix by using Cholesky Decomposition \cite{golub}. In Cholesky Decomposition, we find a lower triangular matrix $L$, such that, the covariance matrix can be written as, 

\subsection{Algorithm for generating nSI CMB temperature realizations}\label{alg}
\begin{figure}[H]
\centering
\subfigure[]{
\includegraphics[width=5.0in,keepaspectratio=true]{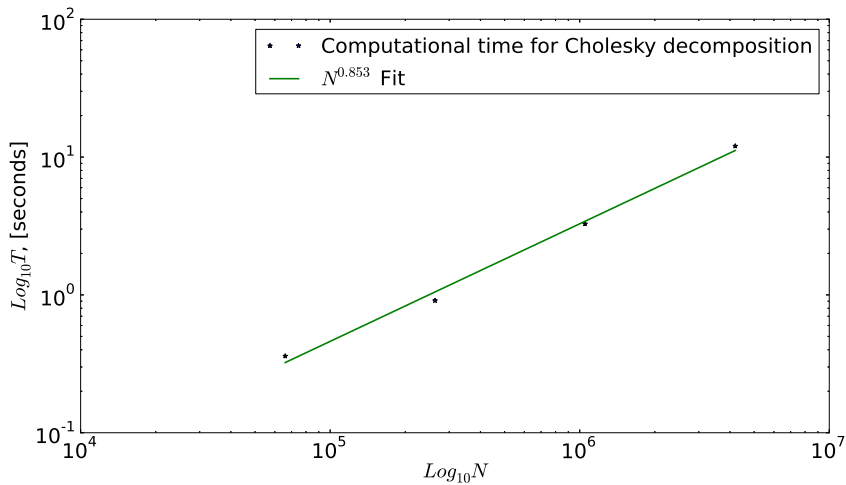}
\label{figtsub1}
}
\subfigure[]{
\includegraphics[width=5.0in,keepaspectratio=true]{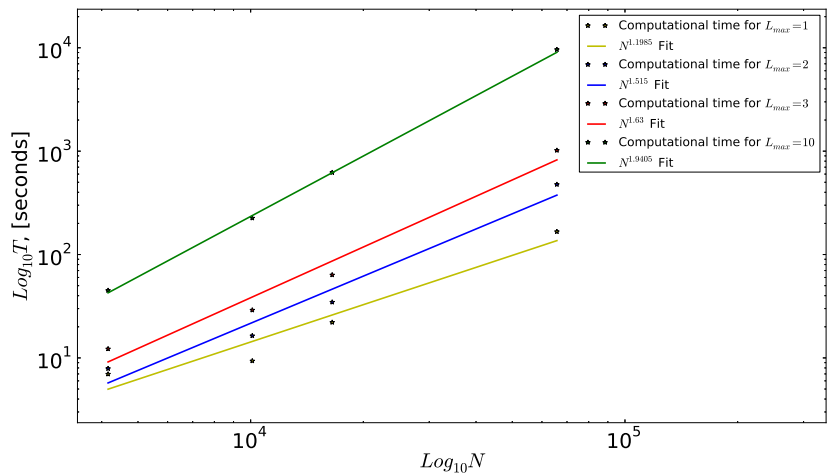}
\label{figtsub2}
}
\caption{$(a)$ Scaling of computational time for Cholesky Decomposition (CD) with $L=1, M=0$ or $L=2, M=0$. Blue circles are the time taken by the code for different $N= l^2_{max}$ values. For $l_{max} =2048$,  computational time requires on a single processor of clock speed $2.60$ GHz  is $\sim 12$ seconds for CD. Green line is  fit to these circles with $N^{0.853} \sim l_{max}^{1.70}$ for CD.
$(b)$ We plot CD for different set of $L_{max}= 1\, (yellow),\, 2\,(blue),\, 3\, (red),\, 10\, (green)$ and $M= [-L_{max}, L_{max}]$ values with  the  dimension of  the SH space covariance matrix, $N$. The scaling of each cases are also shown by a fit in solid line.}\label{figt1}
\end{figure}

\begin{figure}[H]
\centering
\includegraphics[width=5.0in,keepaspectratio=true]{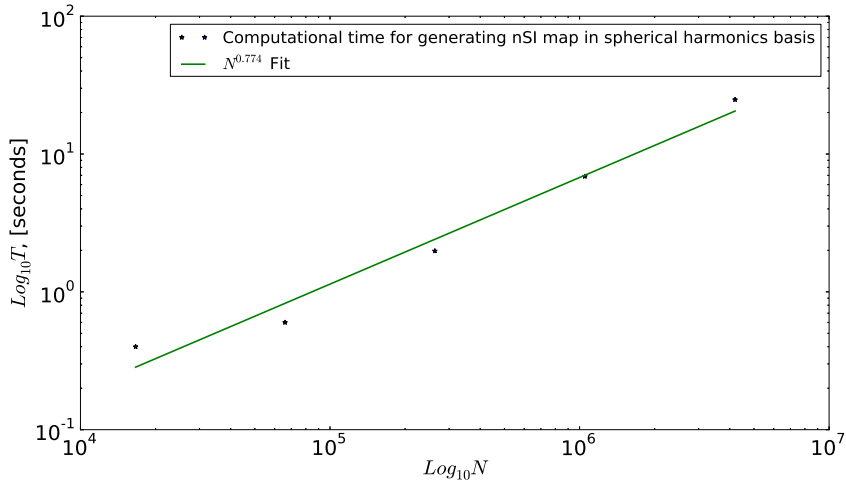}
\caption{Scaling of computational time on a single processor of clock speed $2.60$ GHz for generating a single SH space map, $a_{i}$ as a function of the dimension of  SH space covariance matrix, $N \approx l_{max}^2$.  Blue circles are the time taken by the code for different values of $l_{max}$. For $l_{max} =2048$, with specific BipoSH coefficients, $L=1$ or $L=2$ have a computational time of $\sim 25$ seconds on a single processor of clock speed $2.60$ GHz for generating the maps. Green line is  fit to these circles with $N^{0.77} \sim l_{max}^{1.54}$ for generating nSI maps in SH space.}\label{figcd2}
\end{figure}

In the previous section we discussed the key idea of making nSI simulations of CMB temperature field. In this section, we discuss the logical steps, we use to produce these nSI maps.\\
To efficiently diagonalize the covariance matrix of size $N\approx l_{max}^2$, we use the properties of CG coefficients mentioned in eq.\eqref{eq6}. For a given value of angular power spectra $C_l$ and BipoSH spectra $\alpha^{LM}_{ll'}$, we develop a numerical code to diagonalize the covariance matrix $G$, into lower triangular matrix $\mathcal{L}$ and $\mathcal{L}^\dagger$. $\mathcal{L}$ is also a sparsely populated matrix when incorporated physical effects like scale dependent modulation, Doppler boost and quadrupolar anisotropies. The number of non-zero elements in $\mathcal{L}$ depends upon the number of elements which satisfy the properties of CG coefficients given in eq.\eqref{eq6}. For a dipole statistical anisotropy ($L=1$ non-zero BipoSH coefficients) and a quadrupolar statistical anisotropy ($L=2$ non-zero BipoSH coefficients), only correlated off-diagonal terms are $(l, l+1)$ and $(l, l+2)$ respectively. This makes the covariance matrix sparsely populated. By the choice of coordinate system such that only $M=0$ BipoSH coefficients are non-zero, we can make the matrix very sparse with only one off-diagonal term present in the covariance matrix along with the diagonal term. After CD, nSI maps can be rotated to any other coordinate system, by using the $rotate\_alm$ subroutine of HEALPix \cite{healpix}.\\
 In Fig.~\ref{figtsub1}, we plot the computational time requires for the diagonalization of the covariance matrix for Doppler boost, dipole modulation and quadrupolar anisotropies using CD. Using the properties of CG coefficients, CoNIGS scales with size of the matrix $N$ by $N^{0.853}$ implying  $\sim l_{max}^{1.70}$ for diagonalization. For a denser covariance matrix with more BipoSH multipoles ($L_{max}$), the scaling of the computational time increases. In Fig. \ref{figtsub2}, we also plot the scaling of CD code with $l_{max}$ for different values of $L_{max}$. The computational time for performing the multiplication $\mathcal{L}_{ij}x_j$, to produce real and imaginary part of nSI maps in SH space are plotted in Fig.~\ref{figcd2}. For a given angular power spectra $C_l$ and BipoSH  spectra $\alpha^{LM}_{ll'}$, we need to generate the lower triangular matrix $\mathcal{L}$ only once, and different realizations can be obtained by multiplying $\mathcal{L}$ with different unit variance Gaussian realizations $x_i$. \\
The temperature realizations $a_i$ in SH space are complex numbers with both real and imaginary part, i.e.
\begin{align}\label{eqx1}
a_{i}= c_{i}+ i\,d_{i}, 
\end{align}
where, $c_{i}$ and $d_{i}$ are the real Gaussian random variables. But for the azimuthal symmetric modes $(m=0)$, imaginary part $d_i$ vanishes.  Hence, the map $x_i$ in the new basis should be transformed both to the real part $c_i$  and imaginary part $d_i$ of the temperature map $a_i$. We define $x^R_i$ and $x^I_i$, as two parts of the map $x_i$, which contribute respectively to the real and imaginary part of the map $a_i$, in SH space. Then, the variance of $x_i$ should satisfies,  
\begin{align}\label{eqx2}
\begin{split}
&\sigma^2_{x^R} + \sigma^2_{x^I} \equiv \sigma^2_{x}=1; \hspace{1cm} \text{for $m \neq 0$ modes}\\
&\sigma^2_{x^R} \equiv \sigma^2_{x}=1; \hspace{2cm} \text{for $m = 0$ modes}.
\end{split}
\end{align}
Following are the steps incorporated to produce nSI realizations of CMB temperature field using CoNIGS.
\begin{enumerate}
\item
Compute the covariance matrix using CG subroutine and implement CD on the covariance matrix $G$,  to produce lower triangular matrix $\mathcal{L}$ and its conjugate transpose $\mathcal{L}^\dagger$. This step scales as $l_{max}^{1.7}$, plotted in Fig. \ref{figtsub1}, and takes CPU times $\approx 12$ seconds for $l_{max}= 2048$ on a single processor with $2.6 $ GHz clock speed for Doppler boost, \emph{scale dependent} modulation and quadrupolar anisotropy cases. This step needs to be performed only once for a given covariance matrix.
\item
Generate Gaussian random variables $x^R_i$ and $x^I_i$ with variance satisfying eq.\eqref{eqx2}. This is the temperature map $x_i$, in the new vector space.
\item
Multiply $x_i$ with the lower triangular matrix $\mathcal{L}$, to obtain temperature map, $a_i$ in SH space. This step takes $\approx 25$ seconds of CPU time for producing one SI violated realization in SH space which can incorporate Doppler boost, \emph{scale dependent} modulation and quadrupolar effect for $l_{max}= 2048$. The scaling of the computational time with $l_{max}$ is plotted in Fig. \ref{figcd2}. This step needs to be implemented on every unit variance map $x_i$ to produce many realizations. 
\item
Using the $alm2map$ subroutine in HEALPix \cite{healpix}, we generate the temperature map from $a_{i}$. The scaling of this part is the usual scaling for $alm2map$ subroutine.
\item
Using the HEALPix subroutine $rotate\_alm$, we rotate the nSI maps to change the co-ordinate system.
\end{enumerate}
%All these steps are explained diagrammatically in the flowchart given in Appendix \ref{Flowchart}.\\
%By using the properties of CG coefficients, the computational time required for the code is much less than the usual CD. CD generally scales with dimension of matrix, $N_{max}$ by $N^3$. For making nSI maps, the computational cost depends on $N_{max}$ by $N_{max}^{1.8}$, Fig. \ref{figt1}.

\section{nSI realizations of CMB temperature field}
We study three different cases of nSI maps with the BipoSH spectra that have come under discussions and study due to results obtained by recent experiments like WMAP and Planck. 
\begin{enumerate} 
\item
The quadrupolar $(L=2)$ BipoSH spectra as measured by WMAP \cite{ben}.\\
\item
BipoSH spectra for $L=1$ with a \emph{scale dependent} dipole modulation strength which results in dipolar asymmetry as detected by Planck \cite{Planck}.\\
\item
Doppler boosted CMB temperature map with non-zero BipoSH spectra for $L=1$ as measured by  Planck \cite{Planck_dop}.
\end{enumerate} 
\subsection{Quadrupolar asymmetry: WMAP 7 year measurement of non-zero  BipoSH spectra for $L=2$.}
\begin{figure}[H]
\centering
\includegraphics[width=5.0in,keepaspectratio=true]{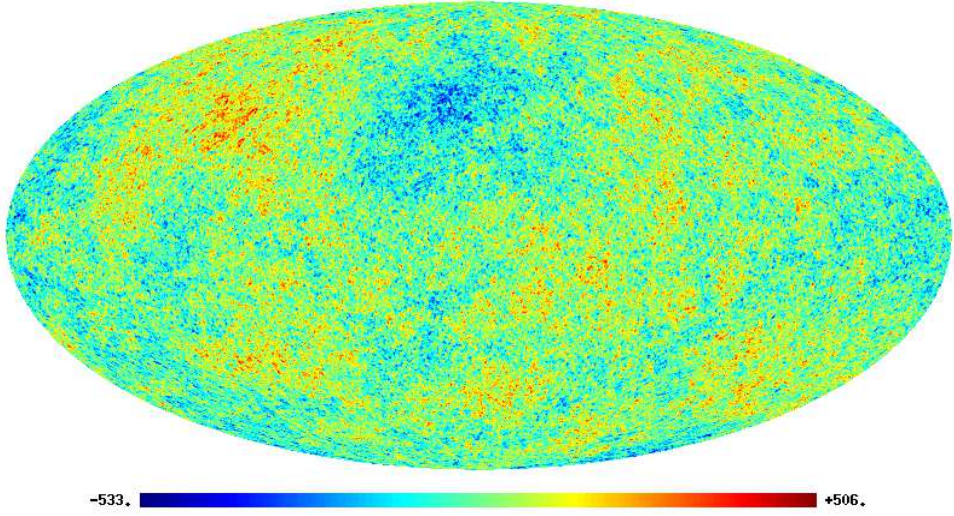}
\caption{nSI realization for CMB  temperature field with $L=2$ BipoSH spectra produced using CoNIGS in ecliptic coordinates.}\label{fig2a}
\end{figure}

\begin{figure}[H]
\centering
%\hspace{1in}
\includegraphics[width=5.0in,keepaspectratio=true]{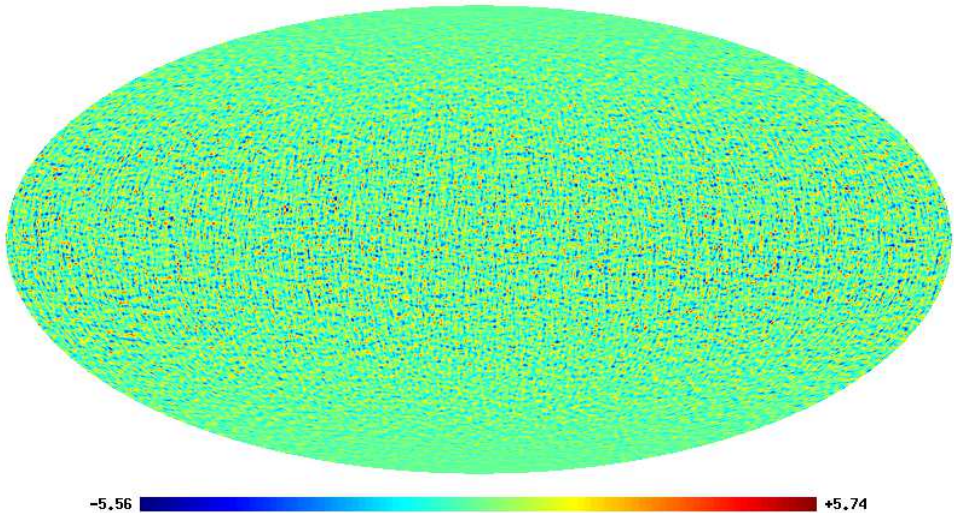}
\caption{Difference between the nSI map with $L=2$ BipoSH spectra and SI map for the same seed of random realization in ecliptic coordinates. The difference in temperature is in the range $[-5.56 \mu K, 5.74 \mu K]$ to be compared to the range of nSI map $[ -534 \mu K,505 \mu K]$ given in Fig.~\ref{fig2a}.}\label{fig2}
\end{figure}

\begin{figure}[H]
\centering
\includegraphics[width=4.2in,keepaspectratio=true]{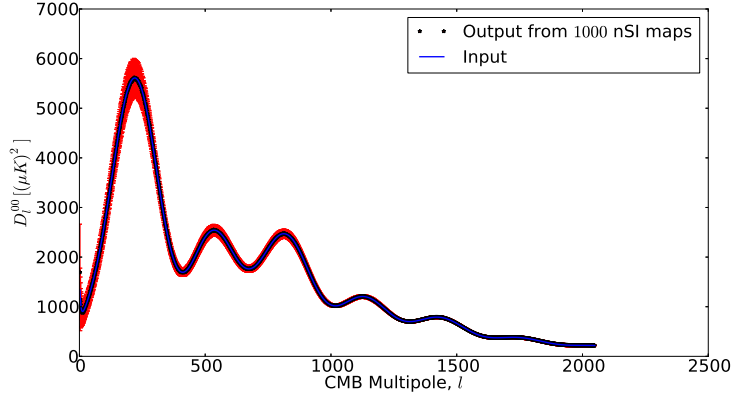}
\caption{Comparison of input and output values of $D^{00}_l= l(l+1)C_l/2 \pi$ obtained from 1000 nSI realizations.}\label{fig:f1}
\end{figure}

\begin{figure}[H]
\centering
\subfigure[]{
\includegraphics[width=4.2in,keepaspectratio=true]{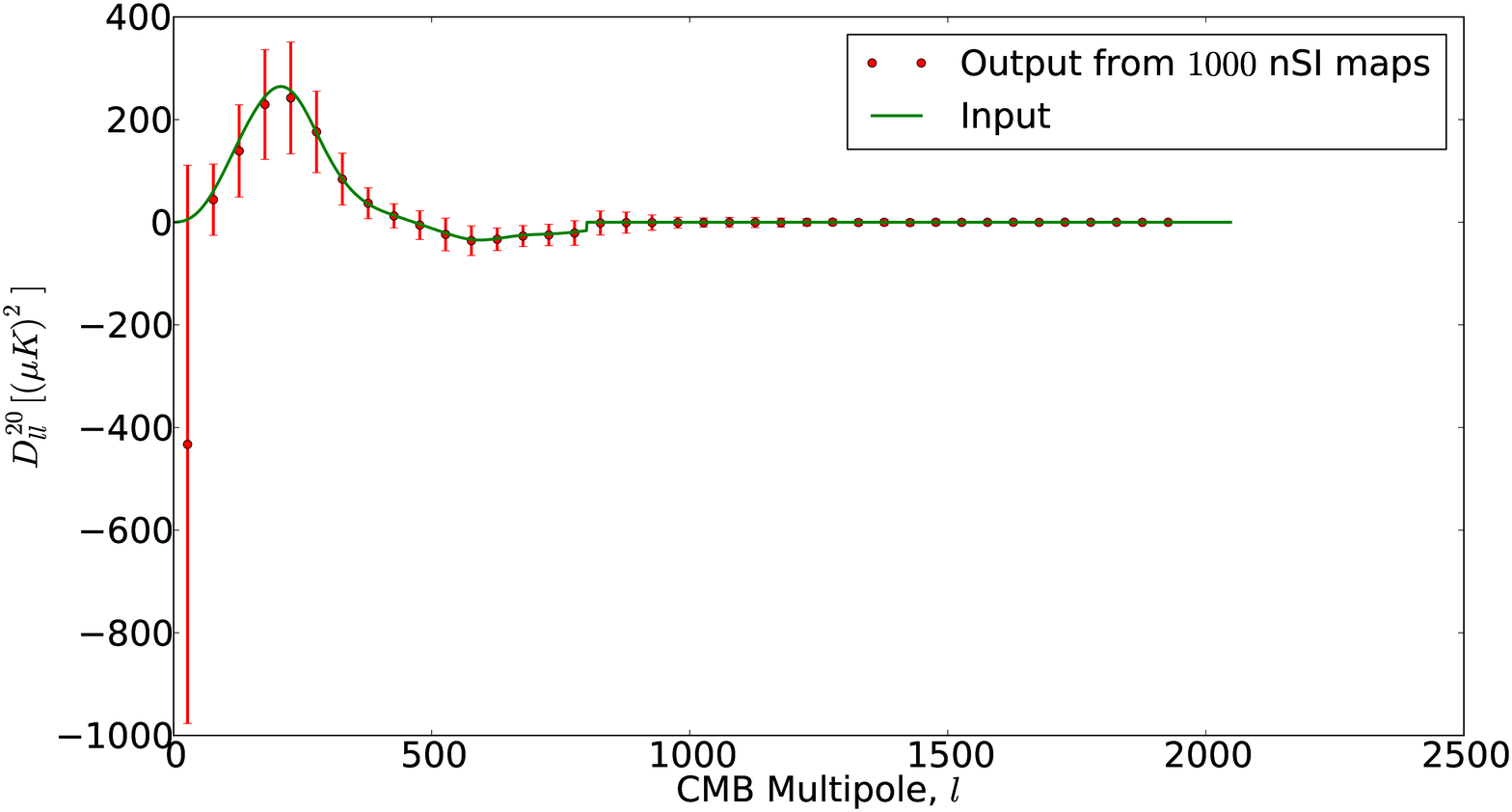}
\label{figsub3}
}
%\caption{Comparison of input and output of BipoSH values $A_{ll}^{20}$ obtained from 100 Non-SI maps with a bin of $\Delta l=20$}\label{fig:f}
%\end{figure}
%\begin{figure}[H]
\centering
\subfigure[]{
\includegraphics[width=4.2in,keepaspectratio=true]{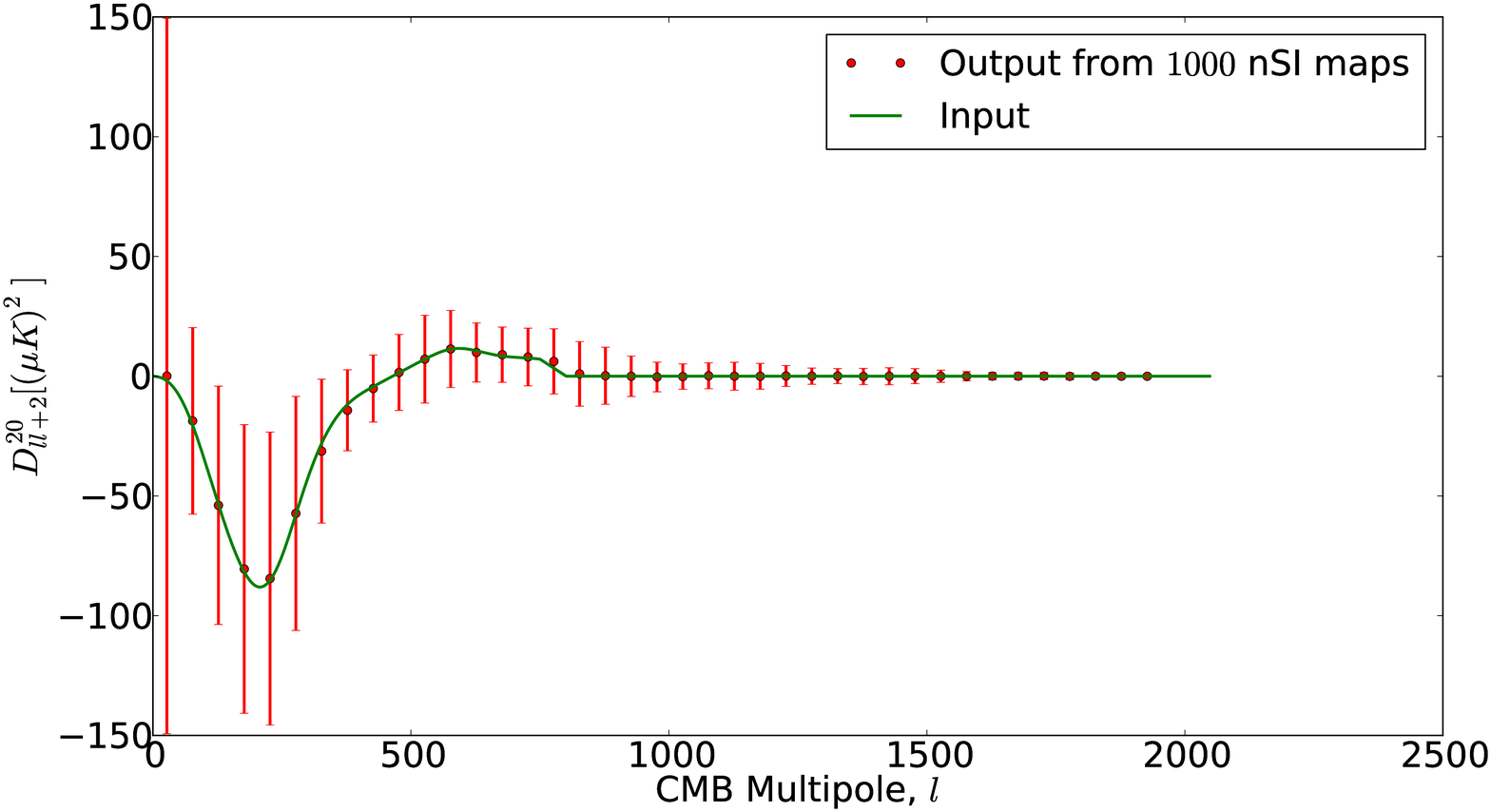}
\label{figsub4}
}
\caption{Comparison of BipoSH spectra, (a) $D^{20}_{ll}= l(l+1)\alpha^{20}_{ll}/2\pi$, (b) $D^{20}_{ll+2}= l(l+1)\alpha^{20}_{ll+2}/2\pi$ obtained from 1000 realizations with the input value of $D^{20}_{ll}$ and $D_{ll+2}^{20}$. Here the BipoSH spectra are binned with $\Delta l =50$. }\label{fig3}
\end{figure}

The measurement of  BipoSH spectra for $L=2, M=0$  by WMAP \cite{ben} is a signature of SI violation. With angular power spectrum $C_l$ and only non-zero BipoSH spectra $\alpha^{20}_{ll}$ and $\alpha^{20}_{ll+2}$, we obtain the covariance matrix $G$ using eq.\eqref{eqbi3c}. Using the numerical algorithm CoNIGS, we obtain the nSI realization of CMB temperature field for $l_{max}= 2048$, given in Fig.~\ref{fig2a}. To show the visual effect of the non-zero BipoSH coefficients, we take the difference of SI and nSI realization (with the same seed value) and is plotted in Fig.~\ref{fig2}. The range of the variation of  difference in temperature fluctuation is  $[-5.56 \rm{\mu K}, 5.74 \rm{\mu K}]$, which is $\sim 100$  times smaller than the range of temperature fluctuation of nSI map. \\
The average two-point correlation function estimated from these maps should match the input covariance matrix $G$.  To test the consistency of the  two-point correlation  for these nSI maps with the input covariance matrix, we obtain $1000$ realizations of nSI maps and then using the BipoSH estimation code \cite{nidhi}, we obtain the two-point correlation from the simulated temperature maps. The comparison between input and output values of  angular power spectra and BipoSH spectra are plotted in Fig.~\ref{fig:f1} and Fig.~\ref{fig3} respectively. 
The comparison between input and output power spectra of temperature field ensures  that the ensemble average of nSI realizations generated by our method  recover the input angular power spectra and BipoSH spectra. This provides an efficient means for creating Monte Carlo ensembles of CMB maps useful in studying different models of SI violation and to study their statistical properties.
\subsection{Dipolar asymmetry: Scale dependent non-zero dipolar $(L=1)$ BipoSH spectra from Planck measurement.}
The nSI estimator on Planck CMB map measured a $3.7 \sigma$ detection of BipoSH spectra for $L=1$, modeled as a dipolar modulation of SI temperature field with a \emph{scale dependent} modulation strength \cite{Planck}. So, it is interesting to make realizations of nSI maps that incorporate this feature.
%Making such nSI realizations for a  scale independent modulation strength can  easily be obtained by multiplying  an SI realization with a given scale independent  modulation function.  However, this procedure cannot be used to make nSI realization for a  \emph{scale dependent} modulation strength. 
CoNIGS is an efficient, fast method to produce nSI realization for  any kind of \emph{scale dependent} modulation field\footnote[1]{nSI realization with scale dependent  modulation can also be made by modulating each temperature multipole as, $T(\hat n)= \sum_l (1+w_l(\hat n))\sum_{m} a_{lm} Y_{lm}(\hat n)$. But the BipoSH spectra for them are slightly different from the BipoSH spectra given in eq.\eqref{eqd4}.}.
%, which is not possible to produce efficiently for any arbitrary modulation case. 
In CoNIGS, we encapsulate \emph{scale dependent} modulation in terms of $L=1$ BipoSH coefficients, which are the complete representation for any SI violation. In terms of BipoSH spectra, \emph{scale dependent} modulation strength can be incorporated by the expression,
\begin{align}\label{eqd4}
\begin{split}
\alpha^{10}_{ll+1}= m^{1}_{l} \bigg[C^{TT}_{l} +C^{TT}_{l+1}\bigg], %\frac{\Pi_{ll+1}}{\sqrt{4\pi}\Pi_{1}}C^{1 0}_{l\, 0\, l+1\, 0},
\end{split}
\end{align}
where, %, $\Pi_{l_1l_2l_3\hdots l_n}= \sqrt{(2l_1+1)(2l_2+1)\hdots (2l_n+1)}$ and 
we used Planck best fit $\Lambda$CDM lensed $C_l$ \cite{Planck4, Planck_param}.
 % Results from Planck \cite{Planck}, consistent with WMAP and COBE  implies violation of SI in the CMB sky.  
% In our method of producing nSI realizations, we can easily incorporate the \emph{scale dependent} modulation strength in terms of the $L=1$ BipoSH coefficients, and for any arbitrary choice of modulation function, our numerical code, CoNIGS can produce the corresponding nSI realizations.
\begin{figure}[H]
\centering
\includegraphics[width=5.0in,keepaspectratio=true]{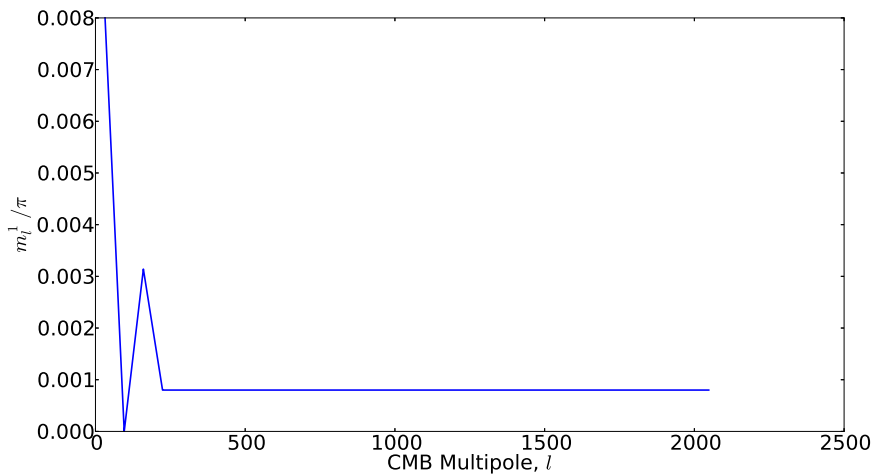}
\caption{Modulation strength $m^1_l$ with a scale dependence as measured by Planck \cite{Planck} used for producing nSI maps.}\label{fig:mode}
\end{figure}
Using CoNIGS, we make realizations of such nSI CMB sky with the BipoSH spectra for $L=1$ with a \emph{scale dependent} modulation strength, $m^1_l$,  given in Fig.~\ref{fig:mode}. This complicated modulation strength is taken to imitate the modulation strength as measured by Planck \cite{Planck}. In Fig.~\ref{fig:d}, we plot a nSI realization  and in Fig.~\ref{fig5} we plot the corresponding difference between SI and nSI realization (with the same seed value) for \emph{scale dependent} dipole modulation. \\Using CoNIGS, we also make realizations with \emph{scale independent} modulation strength with $m_1 =0.008$. The difference between SI map and \emph{scale independent} modulated nSI map is plotted in Fig.~\ref{fig5a}. On comparing the difference maps plotted in Fig.~\ref{fig5} and Fig.~\ref{fig5a}, we can conclude that the \emph{scale independent} modulation strength results in fluctuation at all scales, in contrast to the \emph{scale dependent} modulation, where more fluctuation are at larger angular scale. The comparison between input and output values of angular power spectra and BipoSH spectra from $1000$ \emph{scale dependent} dipole modulated realizations are plotted in Fig.~\ref{fig:f1} and Fig.~\ref{fig6} respectively.

\begin{figure}[H]
\centering
\includegraphics[width=5.0in,keepaspectratio=true]{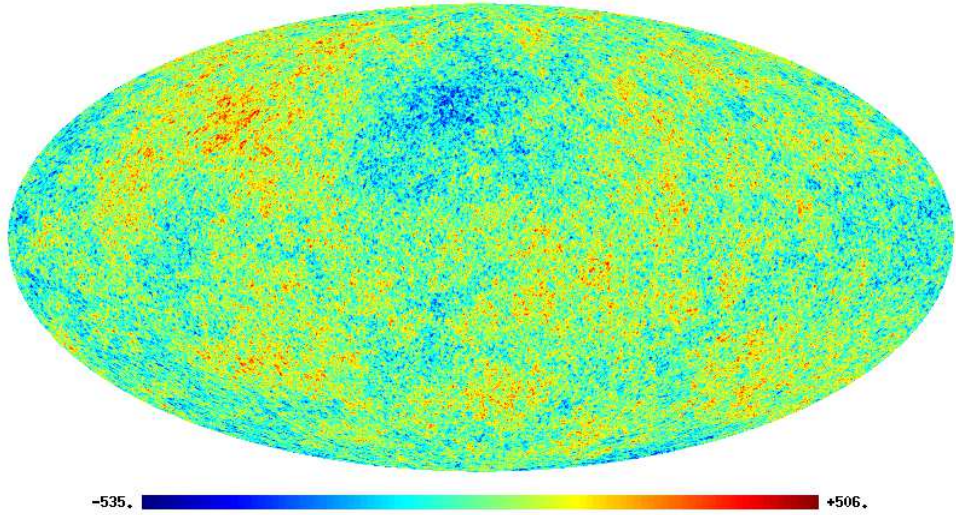}
\caption{nSI realization with $L=1$ non-zero BipoSH spectra due to dipolar asymmetry with a \emph{scale dependent} modulation strength (Fig.~\ref{fig:mode}) as measured by Planck \cite{Planck}.}\label{fig:d}
\end{figure}
\begin{figure}[H]
\centering
\includegraphics[width=5.0in,keepaspectratio=true]{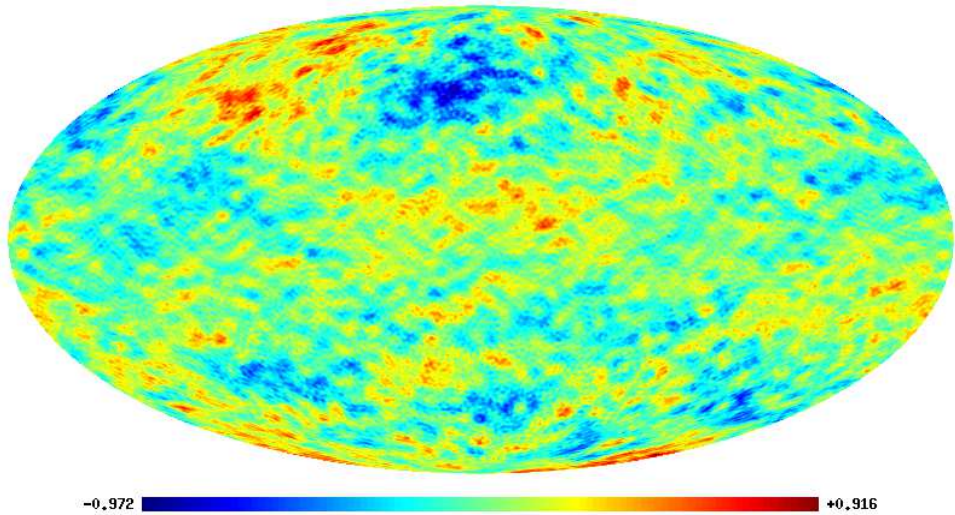}
\caption{Difference between the SI and nSI realization with $L=1$ BipoSH spectra for \emph{scale dependent} modulation strength (Fig.~\ref{fig:mode}) as measured by Planck \cite{Planck}. The difference in temperature is in the range $[-0.972 \mu K, 0.916 \mu K]$ about $\sim 500$ times smaller than nSI map given in Fig.~\ref{fig:d}.}\label{fig5}
\end{figure}
\begin{figure}[H]
%\centering
%\includegraphics[width=5.0in,keepaspectratio=true]{./Planck_map_L_1_modulation_2048_748.eps}
%\includegraphics[width=4.5in,keepaspectratio=true]{./SH_L2_M0.png}
%\caption{nSI map produced by CoNIGS with $L=1$ non-zero BipoSH spectra due to dipolar asymmetry with modulation strength given in Fig.~\ref{fig:mode} as measured by Planck \cite{Planck}.}\label{fig5b}
\centering
\includegraphics[width=5.0in,keepaspectratio=true]{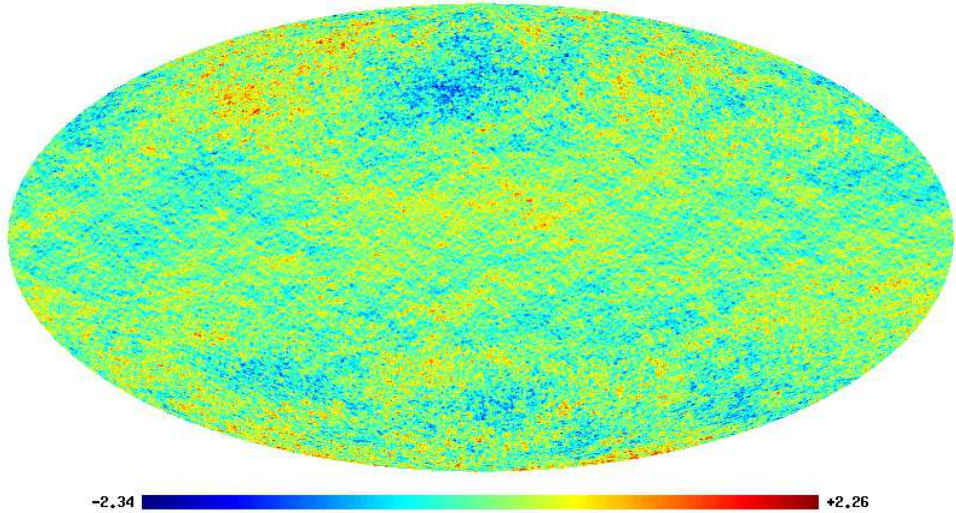}
\caption{Difference between the SI and nSI realization with $L=1$ BipoSH spectra for \emph{scale independent} modulation strength $m_1= 0.008$. The difference in temperature is in the range $[-2.34 \mu K, 2.26 \mu K]$ about $\sim 250$ times smaller than the nSI realization.}\label{fig5a}
%The difference in temperature is in the range $[- 2.24\mu K, 2.36 \mu K]$ about $\sim 250$ times smaller than nSI map given in Fig.\ref{fig2si}.}\label{fig5a}
\end{figure}
\begin{figure}[H]
\centering
\includegraphics[width=5.0in,keepaspectratio=true]{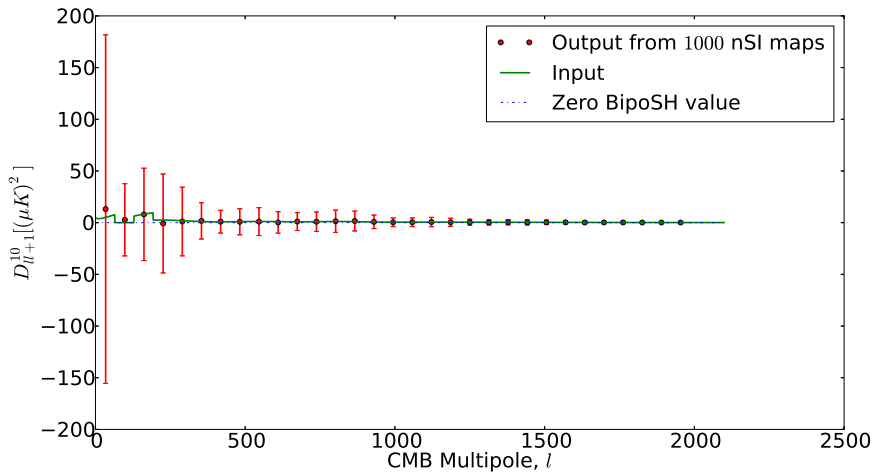}
\caption{Comparison of BipoSH spectra, $D^{10}_{ll+1}= l(l+1)\alpha^{10}_{ll+1}/2\pi$ obtained from  $1000$ realizations of \emph{scale dependent} dipolar asymmetric nSI maps with the input value of  $D^{10}_{ll+1}$. Here, the BipoSH spectra are binned with $\Delta l =64$. The error bar for the $1^{st}$ bin is $168$. The blue line plots zero BipoSH spectra. Using minimum variance estimator \cite{hu2, duncan}, a significant detection of BipoSH spectra is made by Planck \cite{Planck}.}\label{fig6}

%\centering
%\includegraphics[width=6.0in,keepaspectratio=true]{planck_C_l_input_512mod.eps}
%\caption{Comparison of input and output values of $D_l= l(l+1)C_k/2 \pi$ obtained from 100 Non-SI maps}\label{fig:f1}
%\includegraphics[width=6.0in,keepaspectratio=true]{./A1M0D1_comp.eps}
%\caption{Comparison of input and output of BipoSH values $A_{ll}^{20}$ obtained from 100 Non-SI maps with a bin of $\Delta l=20$}\label{fig:f}

%\includegraphics[width=4.5in,keepaspectratio=true]{./allplus219dec512new.eps}
%\caption{Comparison of BipoSH spectra, $D^{10}_{ll+1}= l(l+1)\alpha^{10}_{ll+1}/2\pi$ obtained from  dipolar asymmetric 1000 realizations of CMB temperature sky with the input value of  $D^{10}_{ll+1}$ for a scale independent, constant value of $m_1= 0.008$ . Here the BipoSH spectra are binned with $\Delta l =128$.}\label{fig6}

\end{figure}

\subsection{Doppler boost: Non-zero dipolar $(L=1)$ BipoSH spectra from our local motion.}
\begin{figure}[H]
\centering
\includegraphics[width=5.0in,keepaspectratio=true]{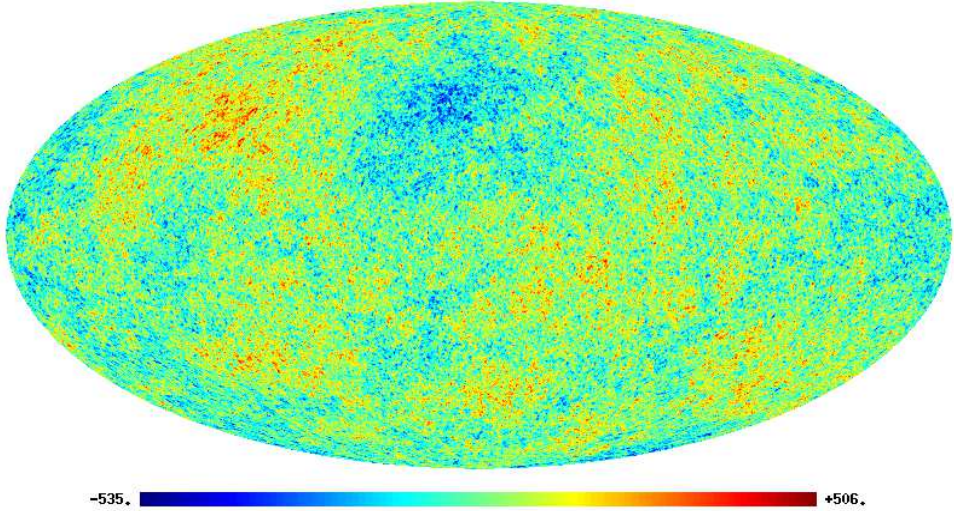}
\caption{nSI realization with $L=1$ non-zero BipoSH spectra due to Doppler boost for $\beta = 1.23 \times 10^{-3}$ along $z$-axis.}\label{fig:e}
\end{figure}

\begin{figure}[H]
\centering
\includegraphics[width=5.0in,keepaspectratio=true]{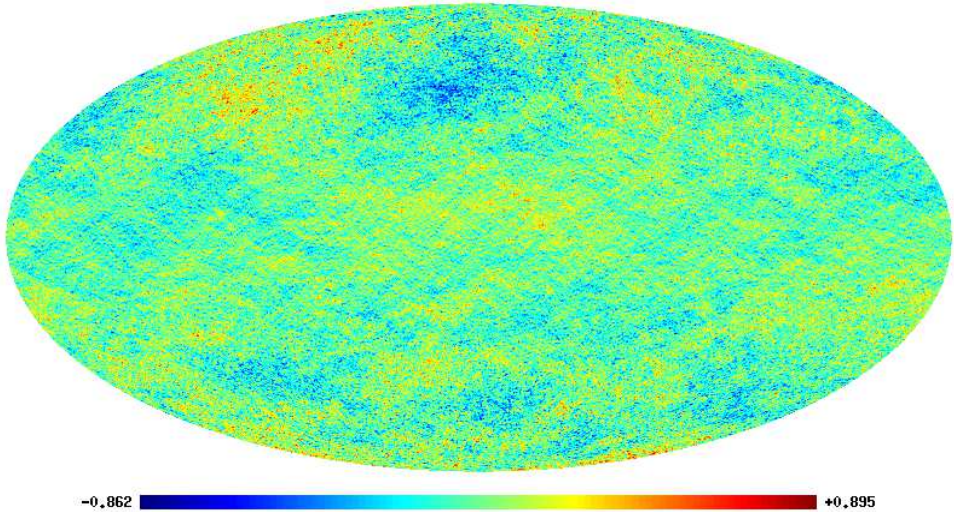}
\caption{Difference between SI and Doppler boosted nSI realization. The difference in temperature is in the range $[-0.862 \mu K, 0.895 \mu K]$ which  is $\sim500$ times smaller than range of fluctuation in the corresponding nSI map given in Fig.~\ref{fig:e}.}\label{fig7}
\end{figure}

\begin{figure}[H]
\centering
%\includegraphics[width=6.0in,keepaspectratio=true]{planck_C_l_input_512mod.eps}
%\caption{Comparison of input and output values of $D_l= l(l+1)C_k/2 \pi$ obtained from 100 Non-SI maps}\label{fig:f1}
\includegraphics[width=5.0in,keepaspectratio=true]{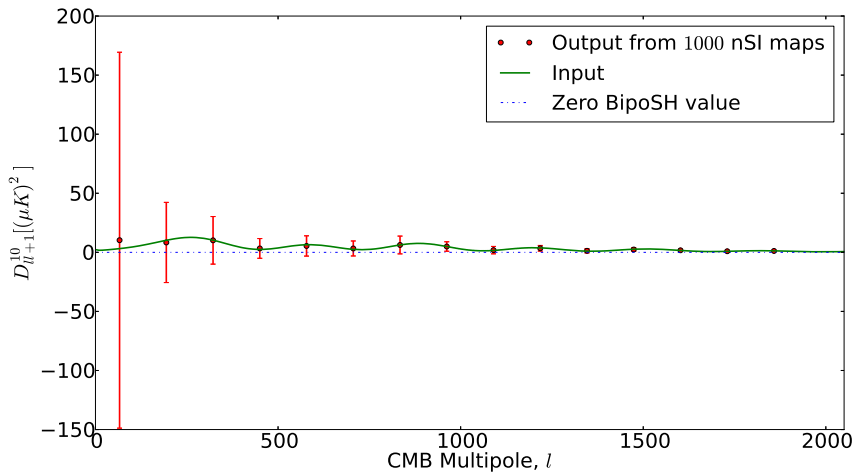}
%\caption{Comparison of input and output of BipoSH values $A_{ll}^{20}$ obtained from 100 Non-SI maps with a bin of $\Delta l=20$}\label{fig:f}

%\includegraphics[width=4.5in,keepaspectratio=true]{./allplus219dec512new.eps}
\caption{Comparison of BipoSH spectra, $D^{10}_{ll+1}= l(l+1)\alpha^{10}_{ll+1}/2\pi$ obtained from  Doppler boosted 1000 realizations of CMB temperature sky with the input value of  $D^{10}_{ll+1}$. Here, the BipoSH spectra are binned with $\Delta l =128$. The error bar for the $1^{st}$ bin is $160$. The blue line plots zero BipoSH spectra. The deviation in Doppler boosted BipoSH spectra from the blue line indicates the possibility for detection of $\beta$ from the small angular scales of CMB temperature field. Using quadratic estimators a significant detection of $\beta$ is made by Planck \cite{Planck_dop}.}\label{fig8}

\end{figure}

Doppler boost of CMB temperature and polarization field due to our local motion with velocity $(\beta \equiv |v|/c = 1.23 \times 10^{-3})$ induces non-zero dipolar $(L=1)$ BipoSH coefficients as shown by Mukherjee et al. \cite{Suvodip}. Recent result from Planck \cite{Planck_dop} estimated $\beta$ from the off-diagonal terms of the SH space covariance matrix which are related to the $L=1$ BipoSH coefficients. Doppler boosting of CMB temperature field results into two kinds of effect, modulation and aberration. Our method is an efficient way of simulating high resolution maps as mentioned in Fig. \ref{figtsub1},\,\ref{figcd2} and also can incorporate aberration and modulation effect in terms of BipoSH spectra up to any order in $\beta$. The linear order effect of  Doppler boost on CMB temperature field leads to non-zero BipoSH  spectra given by \cite{Planck_dop, Suvodip}
\begin{align}\label{eqd3}
\begin{split}
\alpha^{10}_{ll+1}= \beta_{10} \bigg[(l+ b_\nu)C^{TT}_{l} -(l+2 -b_\nu)C^{TT}_{l+1}\bigg],% \frac{\Pi_{ll+1}}{\sqrt{4\pi}\Pi_{1}}C^{1 0}_{l\, 0\, l+1\, 0},
\end{split}
\end{align}
where, $\beta$ is taken along $z$-axis and $b_{\nu}$ is the frequency dependent effect on Doppler boost given by \cite{Planck_dop},
\begin{align} \label{eq11b}
\begin{split}
b_{\nu}= \frac{\nu}{\nu_0}\mathrm{coth}\bigg(\frac{\nu}{2\nu_0}\bigg) -1,
\end{split}
\end{align}
with $\nu_0 = 57$ GHz. We estimate the BipoSH spectra with $b_\nu \approx 3$ corresponding to $\nu =217$ GHz.\\
We plot nSI realizations of CMB temperature in Fig.~\ref{fig:e} with induced Doppler boost for the value of $\beta = 1.23 \times 10^{-3}$. The difference between SI and nSI realization (with the same seed) is plotted in Fig.~\ref{fig7}. The realizations produced by CoNIGS are manifestly nSI and have non-zero value of BipoSH spectra for $L=1$, which we recover back from the BipoSH estimation code \cite{nidhi}. In Fig.~\ref{fig8}, we compare the consistency of input BipoSH spectra with the BipoSH spectra from $1000$ realizations. Power spectra, $C_l$ from the $1000$ simulations also matches the given input $C_l$ as plotted in Fig.~\ref{fig:f1}.

\section{Conclusion}
Statistical Isotropy (SI) violation of CMB temperature and polarization maps are an unavoidable consequence of weak lensing by large scale structures, Doppler boost due to local motion, non-trivial cosmic topology etc. Many observational systematics like masking, non-circularity of beam etc., can also lead to SI violation of CMB temperature field. Recent experiments like WMAP and Planck measured statistically significant non-zero BipoSH coefficients. A likely explanation of non-zero quadrupolar $(L=2)$ BipoSH detection by WMAP as arising from uncorrected  non-circular beam effect is mentioned by several authors \cite{ben, beam1, beam2, beam3, beam}. Doppler boost due to our local motion leads to $L=1$ BipoSH spectra, which is also measured by Planck \cite{Planck_dop}. Planck also made a $3.7 \sigma$  detection of dipolar ($L=1$) BipoSH spectra apart from Doppler boost, known as the dipolar asymmetry \cite{Planck}, which is beyond the understanding of present SI Cosmological models. This  measurement indicates, CMB temperature field to be  manifestly SI violated. To study these effects and to understand the statistics of Non-SI (nSI) CMB sky, it is important to make nSI realizations of CMB temperature and polarization field.\\
In this paper, we have described an efficient  numerical algorithm, Code for Non-Isotropic Gaussian Sky (CoNIGS) developed for generating nSI  Gaussian realizations of CMB temperature field. nSI CMB temperature field leads to non-zero off-diagonal (BipoSH coefficients) terms in the SH space covariance matrix, in contrast to the SI temperature field that has only non-zero diagonal (angular power spectra) terms. The central idea of the technique is to diagonalize once the covariance matrix using Cholesky Decomposition (CD) into lower triangular matrix $\mathcal{L}$ and $\mathcal{L}^\dagger$ as mentioned in eq.\eqref{eq4} and then producing random nSI realization $a_{i}$, by multiplying the lower triangular matrix $\mathcal{L}$ with a unit variance Gaussian realization $x_{j}$, as discussed in Sec.~\ref{cdbasic}. For $L=1$, $L=2$ SI violation cases, the non-zero elements of the covariance matrix are related to the non-zero Clebsch-Gordan (CG) coefficients as mentioned in eq.\eqref{eqbi3c}, which ensures that the covariance matrix is sparsely populated. So, we can diagonalize the covariance matrix faster than the usual CD. The time taken for diagonalization using CoNIGS depends on the dimension of matrix, $N$ by $N^{0.853}$ implying a $\sim l_{max}^{1.70}$ as given in Fig.~\ref{figtsub1}, in contrast to the usual CD, which depends upon $N$ by $N^{3}/6$. On taking non-zero BipoSH coefficients, $A_{ll'}^{LM}$, with larger $L$ values, covariance matrix becomes denser, which results into more computational time as showed in Fig.~\ref{figtsub2}. As mentioned in Sec.~\ref{cdbasic}, nSI realization in SH space are obtained by multiplying lower triangular matrix $\mathcal{L}$ with the unit variance Gaussian realization $x_j$, this step scales with dimension of covariance matrix by $N^{0.77}$ as plotted in Fig.~\ref{figcd2}. \\ Using this method we have generated nSI realizations for temperature field, plotted in Fig.~\ref{fig2a}, with $L=2$ BipoSH spectra as  measured by WMAP. The difference between the SI and nSI realization (with the same seed) is plotted in Fig.~\ref{fig2}. The comparison between input and output values of  angular power spectra and BipoSH spectra from $1000$ realizations are plotted in Fig.~\ref{fig:f1} and Fig.~\ref{fig3} respectively.\\ Planck \cite{Planck} measurement  of BipoSH coefficients for $L=1$ dipolar asymmetry, shows a \emph{scale dependent} modulation as plotted in Fig.~\ref{fig:mode}. Incorporating SI violation in terms of BipoSH coefficients, we can make nSI realizations for  any \emph{scale dependent} modulation strength using CoNIGS. For the \emph{scale dependent} modulation strength, plotted in Fig.~\ref{fig:mode}, nSI  map and the difference between the SI and nSI map for the same seed value of the random realization are plotted in Fig.~\ref{fig:d} and Fig.~\ref{fig5} respectively.  We have also made nSI realizations for \emph{scale independent} modulation strength and plotted the difference between SI realization and nSI realization for a \emph{scale independent} modulation strength, $m_1 =0.008$ in Fig.~\ref{fig5a}. Fig.~\ref{fig5} and Fig~\ref{fig5a} shows the effect  of a \emph{scale dependent} and \emph{scale independent} modulation strength respectively on a SI CMB temperature field. Presence of prominent fluctuations at large angular scales and absence of fluctuations at smaller angular scales in the difference map in Fig.~\ref{fig5}, are the signature of \emph{scale dependent} modulation strength given in Fig.~\ref{fig:mode}. Whereas, for constant modulation strength, Fig.~\ref{fig5a} shows fluctuations at all angular scales. The comparison between the input and output  angular power spectra is plotted in Fig.~\ref{fig:f1}. The plot for comparison between input and output BipoSH power spectra from $1000$ realizations is given in Fig.~\ref{fig6}. \\ We also made realizations for Doppler boosted CMB temperature map which gives rise to $L=1$ BipoSH spectra. The realization for nSI temperature sky and the difference between SI and nSI realization with the same seed value of the random realization are given in Fig.~\ref{fig:e} and Fig.~\ref{fig7} respectively. The comparison between input and output values of  angular power spectra and BipoSH spectra from $1000$ realizations are plotted in Fig.~\ref{fig:f1} and Fig.~\ref{fig8} respectively.  
CoNIGS is a fast and efficient method for producing nSI realizations of CMB temperature field. This is a very important tool to understand the statistics and also to estimate any signal in the nSI temperature field.
\vspace{1cm}

\textbf{Acknowledgement:}\\
We have used the HPC facility at IUCAA. We also used the HEALPix \cite{healpix} package to make the maps. SM acknowledges Council for Science and Industrial Research (CSIR), India, for the financial support as Senior Research Fellow. SM also thanks Santanu Das and Aditya Rotti for many useful discussions. TS acknowledges support from Swarnajayanti fellowship, DST, India.

%\pagebreak[4]
%\appendix

%\section{Flowchart for Making SI violated realizations of temperateure field} \label{Flowchart}
%\begin{figure}[H]
%\centering
%\includepdf[width=7.5in,keepaspectratio=true]{flowchartmapmaking.pdf}
%\end{figure}
\end{document}